\documentclass[aps,prl,twocolumn,superscriptaddress,showpacs]{revtex4}
\usepackage{latexsym}
\usepackage{amsmath}
\usepackage{amssymb}
\usepackage{amsfonts}
\usepackage{graphicx}

\begin{document}

\title{Analytical  description of anomalous diffusion in living cells}

\author{L. Bruno}

\affiliation{Departamento de F\'{\i}sica, Facultad de Ciencias
Exactas y Naturales,\\ Universidad de Buenos Aires, 1428 Buenos
Aires, Argentina.}
\affiliation{Consejo Nacional de Investigaciones Cient\'{\i}ficas
y T\'{e}cnicas, Argentina.}


\author{M. A. Desp\'osito}

\email[]{mad@df.uba.ar}

\affiliation{Departamento de F\'{\i}sica, Facultad de Ciencias
Exactas y Naturales,\\ Universidad de Buenos Aires, 1428 Buenos
Aires, Argentina.}

\affiliation{Consejo Nacional de Investigaciones Cient\'{\i}ficas
y T\'{e}cnicas, Argentina.}


\begin{abstract}

We propose a stochastic model  for intracellular transport processes associated
with the activity of molecular motors.
This out-of-equilibrium model, based on a generalized Langevin equation,
considers a particle immersed in a viscoelastic environment and simultaneously driven by
an external  random force that models the motors activity.
An analytical expression for  the mean square displacement is derived,  which exhibits  a
subdiffusive to superdiffusive  transition.
We show that the experimentally accessible  statistical properties of the diffusive particle motion can be
reproduced by this model.

\end{abstract}




\maketitle


 The intracellular transport of organelles,
vesicles or large proteins involves molecular motors that allow the fast delivery of cargoes
to their correct destination in the cell.  Molecular motors are proteins able to convert the
 energy from the hydrolysis of ATP in directed motion along the cytoskeleton filaments \cite{MM}.
Examples of cytoskeleton motors are kinesin and myosin-V, which move along cytoskeleton filaments such
as microtubules and F-actin \cite{How}.

Single particle tracking techniques have improved significantly in the last years, allowing
capturing the position of micrometer-sized organelles or beads with nanometer and millisecond
resolution \cite{VaLe}.
 Typically, the mean square displacement (MSD) of the particle is analyzed as a function
of the time lag $\tau$ in order to derive the statistical properties of the
transport of large cargoes within the cell \cite{Snider,BBDL}, and to analyze the viscoelastic
 properties of the intracellular environment \cite{Wai}.

Recent experimental works have shown that the MSD of the particle exhibit different dynamical regimes
on different time scales \cite{Arc}.
It has been observed that in the absence of molecular motors \cite{Cas,Won,Git}, or in the case of
ATP depletion \cite{Gal}
the dynamics is subdiffusive.
On the contrary,
a crossover from subdiffusion
(or normal diffusion) to superdiffusion has been reported in experiments in which molecular
motors are active \cite{Cas,Metz,Bru,Kul,Sal,Gal,Len,Bur}.
In this case, the transduction of chemical energy into mechanical work pushes the cell out of equilibrium
\cite{Miz,Lau,Bur,Wil} which implies that the fluctuation-dissipation theorem (FDT) is no longer valid.
Although it is known that the activity of the molecular motors
plays a determinant role in the observed superdiffusive regime, there is no global
model accounting for the relationship between the motors activity and superdiffusion up to now.


In this Letter we propose a stochastic model that takes into account
the previous facts and enables us to reproduce the main features observed in
trajectories of particles driven by molecular motors in living cells.
For this purpose we  describe the intracellular transport by a generalized Langevin equation (GLE) which includes:
\textbf{(i)} a delayed friction function that accounts for the viscoelastic properties of medium,
\textbf{(ii)} a two terms stochastic force:
 a standard \textit{internal} noise due to thermal activity and an \textit{external} noise due to active or
 facilitated transport mediated by molecular motors, and
 \textbf{(iii)} the contribution of the experimental errors.
We obtain a general expression for the MSD of a particle in a viscoelastic
 environment and in the presence of motor forces which can be used to fit experimental data.
 This approach also enables a quantitative description and characterization of the different
 diffusive  regimes  observed in living cells, as was reported in Ref.\cite{Bru}.


The  spontaneous motion of
a particle immersed  in a viscoelastic  environment
is usually described by the generalized Langevin equation (GLE)
\begin{eqnarray}
m \ddot{X}(t) +  \int_0^t dt' \, \gamma (t-t')\, \dot{X}(t')  =  F(t)    \, , \label{Lang}
\end{eqnarray}
where $X(t)$ is the particle position,  $\gamma(t)$ is the dissipative memory kernel and $F(t)$ is the random force.

The integral term accounts for  the  viscoelastic properties
of the medium, with the possibility of storing energy in
the medium and returning it to the particle with a finite
relaxation time.

To  explicitly include deviation from equilibrium
we  assume that the random force $F(t)$ is  the sum of two  uncorrelated contributions, i.e.
$ F(t)=\xi(t)+\chi(t) $,
being  $ \xi(t) $ the standard internal noise due to thermal activity,
and  $\chi(t)$   an external random force that represents the processes that  give rise to the active transport.

The internal  noise $ \xi(t) $, which is responsible for the passive motion, is a zero-centered and stationary
random force with correlation function
$\langle \xi(t) \xi(t') \rangle = C(|t - t'|)$.
It is related  to the memory kernel  $\gamma(t)$   via the FDT \cite{Zwa}
\begin{eqnarray}
C(t)= \frac{k_B T}{m} \, \gamma(t)
 \, ,\label{tfd}
\end{eqnarray}
where $T$ is the absolute temperature  and $ k_B $ is the Boltzmann
constant.

%
It is now well established that the physical origin of
anomalous diffusion is related to  long-time tail correlations \cite{Wa1,MeK}.
In particular, pure power-law correlation functions
are usually employed to model subdiffusive process \cite{Wa1,powerlaw}.  Then, the noise autocorrelation function $C(t)$ can be
chosen as
\begin{eqnarray}
C(t)=  \frac{C_0}{\Gamma(1-\lambda)}\,\left(\frac{t}{\tau_0}\right)^{-\lambda} \, ,
\label{nint}
\end{eqnarray}
where $ 0<\lambda <1 $, $C_0$ is a proportionality coefficient,
$\tau_0$ is an arbitrary characteristic time and $\Gamma(z) $ is the Gamma function.


In addition to the thermal noise, we consider
an  external contribution originated in the activity of  ATP-powered motors.
This external  force  $\chi(t)$, which  is not related to the
dissipation term, is the responsible for the FDT violation.
In other words, deviation from  equilibrium  is directly related to
the irreversible conversion of chemical energy  from ATP hydrolysis into the particle motion \textit{via}
the activity of  molecular motors \cite{BD2}.
This  activity, which   can
be used to generate effectively diffusive movements by
sequences of active directed movements into random directions,
was recently called \textit{active diffusion}  \cite{Klu} and, as we show below,
is the origin of the transition to a superdiffusive regime.

Assuming   that the network on which the active transport occurs has a random organization,
the random force  $\chi(t)$ is chosen as a zero-centered one.
On the other hand, recent works established that the power spectrum
of the noise generated by molecular motors will
be frequency dependent \cite{Bur,Wil,Gal,Metz}.
It was also established that the autocorrelation function of the total noise $F(t)$ has a  power-law behavior
\cite{Lau,Wil,Gal,Bur}.
Accordingly,  we assume a motors force autocorrelation $\Lambda(|t - t'|)=\langle \chi(t) \chi(t') \rangle$,
where $\Lambda(t)$ is given by
\begin{eqnarray}
\Lambda(t)=  \frac{\Lambda_0 } {\Gamma(1-\alpha)}\,\left(\frac{t}{\tau_0}\right)^{-\alpha} \,
,  \label{corract}
\end{eqnarray}
where $\Lambda_0$ is a proportionality coefficient and $0<\alpha <1 $.

It could be thought that the  range chosen for  $\alpha$ is not
 adequate to reproduce the desired superdiffusive behavior, and it  must be $1< \alpha <2$.
However, in Ref.\cite{Porr} it was established that an external noise with a
power-law autocorrelation function like (\ref{corract})
can lead to a superdiffusive behavior
when $\alpha$ is between 0 and 1.
This result, that has been unnoticed in the literature,
 will be explicitly shown  in this work.
Furthermore, considering that the power spectrum of the motors force autocorrelation is
$\widetilde{\Lambda}(\omega)\sim \omega^{\alpha-1}$,
it can be seen that
the limit $\alpha\rightarrow 1$ corresponds
to a series of instantaneous infinite force pulses (white noise limit)
while $\alpha\rightarrow 0$ corresponds to the
 indefinitely large memory case, i.e., the so-called strong
memory limit \cite{Mok}.
Then, an intermediate value of $\alpha$ should correspond
to a smoothing of discontinuities in instantaneous
force pulses, as suggested  in Refs. \cite{Wil,Bur}.
This agrees with the well accepted  picture of molecular motors moving in a step-like manner on microtubules or
actin filaments \cite{steplike}.


On the other hand, the  motion of  organelles or vesicles
is strongly damped in the intracellular
media \cite{How}. Then, the  typical damping time constant is
too short to be appreciable experimentally and thus  the effect
of inertia  can be neglected in (\ref{Lang}).
In this case, and using the Laplace transform technique,
the formal expression for the displacement  can be written as
\begin{eqnarray}
X(t) &=&  x_0  + \int_0^t dt' G(t-t')
F(t') \, , \label{X}
\end{eqnarray}
where $x_0=X(t=0)$ is the deterministic
initial position  of the particle.
The  relaxation function  $G(t)$ is the  Laplace inversion of
\begin{eqnarray}
\widehat{ G}(s) =  \frac{1}{s\,\widehat{\gamma}(s)} \, ,
 \label{Gs}
\end{eqnarray}
where $\widehat{\gamma}(s)= m \,\widehat{C}(s)/k_B T$ is the Laplace transform of the
dissipative memory kernel.
The relaxation function  (\ref{Gs}) is independent of the external noise and it is
equal to the one  obtained  in the standard internal noise case when inertial effects are neglected.


 Typically, the particle trajectory is quantitatively
analyzed in terms of the mean square displacement (MSD), which is calculated as
$\langle \left(X(t+\tau)-X(t)\right)^2 \rangle$
where $|X(t+\tau)-X(t)|$ is the particle displacement between two time points,
 $t$  denote  the \textit{absolute time} while  $\tau$ is the so-called
\textit{lag time}.

To obtain an analytical expression for  the MSD,   it is necessary  to consider the
two-time correlation dynamics.
Using Eq. (\ref{X})   we can write the displacement two-time correlation  as
\begin{eqnarray}
\langle X(t)X(t') \rangle & =& x^{2}_0+
 \int_0^{t} dt_1
G(t-t_1)\times \nonumber \\
&& \int_0^{t'} dt_2 G(t'-t_2) \langle F(t_1) F(t_2)
\rangle   \, . \label{2tc}
\end{eqnarray}

Since $ F(t)=\xi(t)+\chi(t) $, the integral  containing  the
correlation function $\langle F(t_1) F(t_2)
\rangle $ can be split into the internal and external contributions.
Using relation (\ref{tfd}), and considering the symmetry properties of
the correlation functions $C(t)$ and $\Lambda(t)$,
the two-time position correlation function (\ref{2tc}) can be written as
\begin{eqnarray}
\langle X(t+\tau)X(t) \rangle & =& x_0^{2} +
 k_B T (I(t) + I(t+\tau) - I(\tau))
 \nonumber \\
& + &  \int_0^{t} dt_1
 \left\{ G(t_1) H(t_1+\tau)\right.
  \nonumber \\
&& \qquad \left. + \, G(t_1+\tau) H(t_1) \right\}
\, ,
\label{xx}
\end{eqnarray}
where
\begin{eqnarray}
  I(t)&=&\int_0^t dt' G(t') \, , \\
  H(t)&=&\int_0^t dt' G(t')\Lambda(t-t') \, .
\end{eqnarray}

Note that, while the relaxation functions $G(t)$ and $I(t)$ only depend on the
internal thermal noise through the memory kernel $\gamma(t)$,
the relaxation function $H(t)$  includes the contribution of the external random force.


For the autocorrelation functions  given by  (\ref{nint}) and (\ref{corract})
 the involved relaxation functions
can be written as

\begin{eqnarray}
I(t)  & = & \frac{k_B T}{C_0}\frac{1}{\Gamma(\lambda+1)} \,\left(\frac{t}{\tau_0}\right)^{\lambda} \, , \label{It}\\
G(t)  & = & \frac{k_B T}{\tau_0 \, C_0 }\frac{1}{\Gamma(\lambda)} \,\left(\frac{t}{\tau_0}\right)^{\lambda-1} \, , \label{Gt}\\
H(t) & = & \varepsilon \, k_B T \,  \frac{1}{ \Gamma(\lambda-\alpha+1)} \,\left(\frac{t}{\tau_0}\right)^{\lambda-\alpha}\,   ,\label{Ht}
\end{eqnarray}
where $\varepsilon = \Lambda_0/C_0$ is a dimensionless parameter that measures
the relative intensity among
the motors force and the thermal random force.


Finally, $\langle \left(X(t+\tau)-X(t)\right)^{2} \rangle$ can be calculated using (\ref{xx})
together with  Eqs. (\ref{It}) to (\ref{Ht}).
Even though  the result depends on the relation between $\lambda$ and $\alpha$,
it can be demonstrated  that for $2 \lambda -\alpha > 0$
the MSD have an analytical expression given by
\begin{widetext}
\begin{eqnarray}
MSD(t,\tau) & = & \frac{2k_B T}{\gamma_0}
\frac{1}{\Gamma (\lambda +1)}(\frac{\tau}{\tau_0})^{\lambda }
+  \varepsilon\, \frac{2k_B T}{\gamma_0} \frac{1}{\Gamma (\lambda )\Gamma (\lambda-\alpha+1  ) }  \times \nonumber \\
&& \left\{
\frac{1}{2\lambda-\alpha} \frac{(t+\tau )^{2 \lambda -\alpha }+t^{2 \lambda -\alpha }}{\tau_0^{2 \lambda -\alpha }}
+ (\frac{\tau}{\tau_0})^{2 \lambda -\alpha }
 \left((-1)^{\lambda+\alpha } B_{-\frac{t}{\tau }}(\lambda-\alpha +1,\lambda)- (-1)^{-\lambda } B_{-\frac{t}{\tau }}(\lambda ,\lambda-\alpha+1) \right)
  \right\}
 \nonumber \\
     \label{msdttau}
\end{eqnarray}
\end{widetext}
where $\gamma_0 = C_0/k_B T$ and $B_{x}(a,b )$ is the incomplete beta function \cite{beta}.
While the first term of (\ref{msdttau}) represents the subdiffusive behavior due to thermal activity, the second one
has its origin on the activity of the external random forces.


Note that the MSD  (\ref{msdttau}) is an aging variable depending on the absolute time
$t$ and the time lag $\tau$ \cite{aging}.
However, in typical intracellular tracking experiments an organelle or endosome is  followed during 10-100 seconds.
This time is  much shorter than  the sample preparation durations (absolute time).
Then, it can be considered that the experimental measured MSD  is equivalent to
the long time  limit
\begin{eqnarray}
MSD(\tau)=  \lim_{ t \to \infty } \langle \left(X(t+\tau)-X(t)\right)^{2} \rangle \, .
\label{ltlmsd}
\end{eqnarray}


On the other hand, to make a comparison  with experimental results it is necessary to take into account
measurement errors on the particle position determination intrinsic to the SPT experiment or originated in
 biological activity. It has been established that this effect can be introduced
by adding  an uncorrelated noise of variance
$\eta^2$ to the mean square displacement \cite{errors}.


Then, using the asymptotic expansions for the incomplete beta function \cite{beta}
in (\ref{msdttau}) and including the measurement errors,
  the  MSD  (\ref{ltlmsd}) can be finally written as
\begin{eqnarray}
MSD(\tau) &= &
\frac{2k_B T}{\gamma_0}
\left\{\frac{1}{\Gamma (\lambda +1)}(\frac{\tau}{\tau_0})^{\lambda }
+ \varepsilon  K_{\lambda,\alpha}(\frac{\tau}{\tau_0})^{2 \lambda -\alpha }\right\} \nonumber \\
& & \qquad  + (2\eta)^2  \, ,
\label{msdanomal}
\end{eqnarray}
where
\begin{eqnarray}
K_{\lambda,\alpha}=\Gamma (\alpha -2 \lambda ) \, \frac{\sin (\pi  (\lambda -\alpha ))-\sin (\pi  \lambda )}{\pi }
\, ,
\end{eqnarray}
is a positive constant for $2 \lambda -\alpha>0$ .

It is worth pointing out that  the second term  of (\ref{msdanomal}) is a
superdiffusive contribution to the MSD when  $1<2 \lambda -\alpha <2 $.
In this case, our model predicts a crossover from a  subdiffusive to a superdiffusive regimes,
with exponents $\lambda$ and $2 \lambda -\alpha$, respectively.
This transition can be interpreted as follows: for short enough times  the measurements errors,
represented by $(2\eta)^2$,  dominate,
for intermediate time scales a subdiffusive behavior due to the viscoelastic properties of the intracellular medium prevails,
while at longer time scales motors activity effects dominate leading to a superdiffusive behavior.


The presented model is characterized by four parameters: $\lambda$, $\alpha$, $\varepsilon$ and $\eta$,
where $\lambda$ and $\alpha$ are the power law exponents of the internal and external noise
correlation functions, $\varepsilon$ is a parameter that measures the relative intensity
between random forces and $\eta$ is associated with the residual value of the
MSD as $\tau\rightarrow 0$.
Also,  the magnitude of the  force
exerted by the motors can be estimated as  $ F_{mot} \approx \sqrt{\Lambda_0/\Gamma(1-\alpha)}$
 where $\Lambda_0$ can be obtained in terms of the involved parameters \cite{Bru}.


Interestingly, some recent works have used an empirical three parameters model  of
the form $A+ D^{*} t^{ \beta}$, to fit the MSD vs. time lag \cite{Rau,BBDL}.
This approach has been used indistinctly for systems showing subdiffusive ($\beta<1$)
or superdiffusive ($\beta>1$) behaviors. However, as shown above, different  regimes can coexist and
our model allows to describe both situations with a unique set of parameters,  as we show in Ref.\cite{Bru}.
For example, if in Eq.(\ref{msdanomal}) the
noise $(2\eta)^{2} $ dominates over the
subdiffusive term in the measurement temporal range, the  empirical expression with $\beta>1$ holds.
On the other hand, the subdiffusive behavior observed in the absence of molecular motors or ATP depletion can be  reproduced
setting $\varepsilon=0$ in (\ref{msdanomal}).


In conclusion,  we have presented a model that  provides
a physical interpretation of the crossover from subdiffusive to superdiffusive behavior
observed in single particle tracking experiments
in living cells.
A similar approach was recently introduced in the literature \cite{Wil,Gal}.
However, in these works all the forces (internal and external) contributions are
included in a single term and thus, they do not distinguish
between the thermal and the active forces, a key element to determine
motor forces \textit{in vivo}.
We believe that the present approach can be used to analyze any single particle
tracking data set obtained in the observation of
intracellular transport driven by molecular motors in  living cells.
\\

We thank  Valeria Levi for  a careful reading of the manuscript and helpful discussions.
This work was
supported by grants PICT 928/06 and  PICT 31980/05  from Agencia Nacional de Promoci\'{o}n Cient\'{i}fica y
Tecnol\'{o}gica, Argentina.





\begin{thebibliography}{}
\bibitem{BD2} Deviation from the
internal equilibrium situation can quantified by an  effective temperature, which can be
 evaluated  trough the second moment of the velocity:
L.Bruno and M. A. Desp\'{o}sito, to be submmited.
\end{thebibliography}
\end{document}